\documentclass[twocolumn,aps,prd]{revtex4-2}
\usepackage{graphicx}
\usepackage{bm}
\usepackage{relsize}
\usepackage{amsmath,amsfonts,amsthm}

\begin{document}
\newcommand*{\bi}{\bibitem}
\newcommand*{\ea}{\textit{et al.}}
\newcommand*{\eg}{\textit{e.g.}}
\newcommand*{\zpc}[3]{Z.~Phys.~C \textbf{#1}, #2 (#3)}
\newcommand*{\plb}[3]{Phys.~Lett.~B \textbf{#1}, #2 (#3)}
\newcommand*{\phrc}[3]{Phys.~Rev.~C~\textbf{#1}, #2 (#3)}
\newcommand*{\phrd}[3]{Phys.~Rev.~D~\textbf{#1}, #2 (#3)}
\newcommand*{\phrl}[3]{Phys.~Rev.~Lett.~\textbf{#1}, #2 (#3)}
\newcommand*{\pr}[3]{Phys.~Rev.~\textbf{#1}, #2 (#3)}      
\newcommand*{\npa}[3]{Nucl.~Phys.~A \textbf{#1}, #2 (#3)}  
\newcommand*{\npb}[3]{Nucl.~Phys.~B \textbf{#1}, #2 (#3)}  
\newcommand*{\npbps}[3]{Nucl.~Phys.~B (Proc. Suppl.) \textbf{#1}, #2 (#3)}  
\newcommand*{\ptp}[3]{Prog. Theor. Phys. \textbf{#1}, #2 (#3)}
\newcommand*{\ppnp}[3]{Prog. Part. Nucl. Phys. \textbf{#1}, #2 (#3)}
\newcommand*{\epjc}[3]{Eur. Phys. J. C \textbf{#1}, #2 (#3)}
\newcommand*{\jpg}[3]{J. Phys. G \textbf{#1}, #2 (#3)}
\newcommand*{\mpla}[3]{Modern Physics Letters A \textbf{#1}, #2 (#3)}
\newcommand*{\jhep}[3]{J. High Energy Phys. \textbf{#1}, #2 (#3)}
\newcommand*{\rmph}[3]{Rev. Mod. Phys. \textbf{#1}, #2 (#3)}
\newcommand*{\ijmpa}[3]{Int. J. Mod. Phys. A \textbf{#1}, #2 (#3)}
\newcommand*{\cpc}[3]{Chin. Phys. C \textbf{#1}, #2 (#3)}
\newcommand*{\ra}{\rightarrow}
\newcommand*{\pippim}{\pi^+\pi^-}
\newcommand*{\kpkm}{K^+K^-}
\newcommand*{\kskl}{K^0_SK^0_L}
\newcommand*{\rf}[1]{(\ref{#1})}
\newcommand*{\be}{\begin{equation}}
\newcommand*{\ee}{\end{equation}}
\newcommand*{\bea}{\begin{eqnarray}}
\newcommand*{\eea}{\end{eqnarray}}
\newcommand*{\ba}{\begin{array}}
\newcommand*{\ear}{\end{array}}
\newcommand*{\nl}{\nonumber \\}
\newcommand*{\rmd}{\mathrm d}
\newcommand*{\die}{e^+e^-}
\newcommand*{\jj}{\mathrm i}
\newcommand*{\ndf}{\mathrm{NDF}}
\newcommand*{\mev}{\mathrm{~MeV}}
\newcommand*{\cndf}{\chi^2/\mathrm{NDF}}
\newcommand*{\minuit}{\texttt{MINUIT}}
\newcommand*{\reduce}{\texttt{REDUCE}}
\newcommand*{\Mathematica}{\texttt{Mathematica}}
\newcommand*{\w}{\sqrt s}
\newcommand*{\e}[1]{{\mathrm e}^{#1}}
\newcommand*{\ie}{\textrm{i.e.}}
\newcommand*{\dek}[1]{\times10^{#1}}
\newcommand*{\vv}{V \bar V}
\newcommand*{\dd}{D^{\ast+}D^{\ast-}}
\newcommand*{\dds}{D_s^{\ast+}D_s^{\ast-}}
\newcommand*{\dstp}{D^{\ast+}}
\newcommand*{\dstm}{D^{\ast-}}
\newcommand*{\rd}{\mathrm d}
\newcommand*{\p}{$p$-value }

\title{Lagrangian-based model of the $\bm{e^+e^- \rightarrow V \bar V}$
process applied to the $\bm{e^+e^- \rightarrow D^{\ast+} D^{\ast-}}$ data}
\author{Peter Lichard}
\affiliation{
Institute of Physics and Research Centre for Computational Physics
and Data Processing, Silesian University in Opava, 746 01 Opava, 
Czech Republic
}
\begin{abstract}
Using the quantum field theory, we derive a Breit-Wigner-type formula
for the $e^+ e^-$ annihilation into a vector meson and its antiparticle,
and relate the formula parameters to observable quantities. The formula
soundness is checked by fitting the $e^+ e^- \to D^{\ast+}D^{\ast-}$
data published by the BESIII Collaboration in 2022.
\end{abstract}
\date{\today}
\maketitle
\section{Introduction}
In recent years, many experiments on the production of open-charm-meson pairs 
at $\die$ colliders have been performed. Their survey can be found in
\cite{wang}. Among them, experiments of the type not studied with
light mesons appeared, namely, the production of vector-meson pairs. The
newcomers are $\dd$ and $\dds$.

In one of those experiments \cite{besiii2023}, the observed $\die\to\dds$ 
cross section was fitted by the standard Breit-Wigner (BW) formula. 

Recently, we have shown \cite{pl1} that the vector-meson dominance (VMD) 
hypothesis, combined with the Lagrangian formalism of the quantum field theory,
leads to the $\die\to D\bar D$ cross section formula, which is identical to 
the standard BW formula. What concerns the $\die$ annihilation 
into a vector (V) and a pseudoscalar meson (P), the same procedure gives a 
different formula \cite{vps}. Analyzing the latter, we find that at high 
energies it behaves like $s^{-2}$, where $s$ is the CMS energy squared. whereas 
the standard BW formula behaves like $s^{-3}$. It suggests that the BW formula
is not universally valid, but should be replaced by other formulas if
the final state is not of the P-P type.

In this Letter, we first present the derivation of the $\die\to V\bar V$ 
cross-section formula based on the VMD hypothesis and the Lagrangian
formalism of the quantum field theory. Then we explore it by fitting the
BESIII 2022 $\die\to\dd$ data \cite{besiii2022}.
The fit provides good results, with a \p of 52\% and a satisfactory agreement 
with the PDG \cite{pdg2026} $\psi$ charmonia parameters. 

\section{Model}
\label{model}
\begin{figure}[]
\includegraphics[width=0.39\textwidth,height=0.13\textwidth]{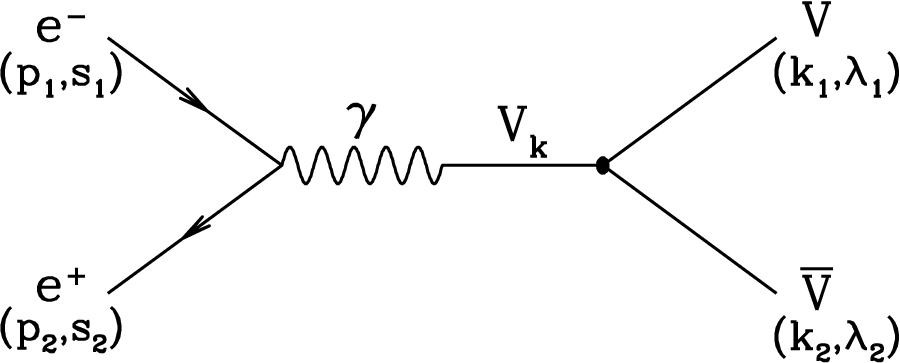}
\caption{\label{fig:ee2vbarv}Feynman diagram defining our model. The $V_k$
represents a truly neutral vector-meson resonance that can couple to the
$\vv$ pair.}
\end{figure}

For the description of the electron-positron annihilation into a $\vv$ pair, 
we are going to build a VMD model based on the Feynman diagram 
depicted in Fig.~\ref{fig:ee2vbarv}
and the interaction Lagrangian
\bea
\label{lagr}
{\cal L}(x)&=&\jj G_{V_kV\bar V}\Big(V_k^\mu(x) V^\dagger_\nu(x)
\stackrel{\leftrightarrow}{\partial_\mu}V^\nu(x)+V^\dagger_\mu(x)\nl
&\times& V_\nu(x)\stackrel{\leftrightarrow}{\partial^\mu}V_k^\nu(x)
+V^\mu(x) V_k^\nu(x)\stackrel{\leftrightarrow}{\partial_\mu}
V^\dagger_\nu(x)\Big),
\eea
where $V(x)$ is the vector field, the quanta of which are mesons $V$ and 
$\bar V$, and the $V_k(x)$ denotes a (hermitian) field pertinent to the
intermediate vector meson $V_k$. The $\gamma V_k$ junction is parametrized as 
$eM_k^2/g_k$ in analogy with the $\gamma\rho^0$ junction 
$eM^2_{\rho^0}/g_\rho$.

The interaction Lagrangian \rf{lagr} implies the three-vector-meson vertex 
that is well known in the Standard Electroweak Model ($ZW^+W^-$) and QCD
(three gluons). Its implication for our 
case, where the incoming $V_k$ meson with four-momentum $q^\mu$ meets with the 
outgoing $V$ ($k_1^{\alpha_1}$) and $\bar V$ ($k_2^{\alpha_2}$), leads,
after applying the four-momentum conservation, to
\bea
\label{vertex}
V^{\mu\alpha_1\alpha_2}(k_1,k_2)&=&g^{\alpha_1\alpha_2}(k_1-k_2)^\mu
+g^{\alpha_2\mu}(k_1+2k_2)^{\alpha_1}\nl
&-&g^{\mu\alpha_1}(k_2+2k_1)^{\alpha_2}.
\eea
Representing all elements from Fig. \ref{fig:ee2vbarv} we are getting the
following amplitude
\bea
\label{amplitude}
{\cal M}&=&-\frac{\jj e^2}{s}\frac{R_k}{s-M_k^2+\jj M_k\Gamma_k}
\bar{v}(p_2,s_2)\gamma_\mu u(p_1,s_1)\nl
&\times&\epsilon^*_{\alpha_1}(k_1,\lambda_1)
\epsilon_{\alpha_2}(k_2,\lambda_2)V^{\mu\alpha_1\alpha_2}(k_1,k_2)
\eea
where $M_k$ and $\Gamma_k$ are the mass and total width of the $V_k$
resonance, respectively, and 
\be
\label{rk}
R_k=M_k^2G_{V_kV\bar V}/g_k.
\ee
It is worth noticing that in this process, the usual momentum-dependent 
term in the numerator of the vector meson propagator does not appear 
in \rf{amplitude}.

The complex conjugate amplitude is
\bea
{\cal M^\ast}&=&\frac{\jj e^2}{s}\frac{R_k}{s-M_k^2-\jj M_k\Gamma_k}
\bar{u}(p_1,s_1)\gamma_\nu v(p_2,s_2)\nl
&\times&\epsilon_{\beta_1}(k_1,\lambda_1)
\epsilon^*_{\beta_2}(k_2,\lambda_2)V^{\nu\beta_1\beta_2}(k_1,k_2)\nonumber.
\eea
When calculating the sum over initial spins and final polarizations
of the amplitude magnitude squared, we use the well-known manipulations 
with spinors (setting $m_e=0$) and the formula for the polarization vector sum
\[
S_{\alpha\beta}(k)=\sum_\lambda \epsilon^*_\alpha(k,\lambda)
\epsilon_\beta(k_,\lambda)=-g_{\alpha\beta}+\frac{k_\alpha k_\beta}{k^2},
\]
to arrive at the formula
\[
\sum_{\ba{c}s_1,s_2\\ \lambda_1,\lambda_2\ear}|{\cal M}|^2=e^4
\left|\frac{R_k}{s-M_k^2+\jj M_k\Gamma_k}\right|^2X,
\]
where 
\bea
X&=&4\left[p_{1\mu}p_{2\nu}+p_{2\mu}p_{1\nu}-(p_1.p_2)g_{\mu\nu}\right]
V^{\mu\alpha_1\alpha_2}(k_1,k_2)\nl
&\times&V^{\nu\beta_1\beta_2}(k_1,k_2)
S_{\alpha_1\beta_1}(k_1)S_{\alpha_2\beta_2}(k_2)/s^2\nonumber.
\eea
Using the algebraic programming system \reduce~of Anthony C. Hearn
\cite{reduce}, we can 
express $X$ as a function of $s$ and $t=(k_1-p_1)^2$
\bea
X&=&\frac{2}{m^4s^2}[-12m^8+4m^6s-17m^4s^2+4m^2s^3 \nl
&+& (24m^6-20m^4s+6m^2s^2-s^3)\,t\nl
&-&(12m^4-4m^2s+s^2)\,t^2],\nonumber
\eea
where $m$ is the $V$ mass. Our final goal is to get the cross 
section from the formula
\[
\sigma=\frac{1}{64\pi s}\frac{1}{|\vec p_1|^2}\int
\overline{|{\cal M}|^2}\,\rd t,
\]
where the overline denotes averaging over initial spin states and summing 
over final polarizations. For that purpose, we calculate
\[
Y=\int_{t_1}^{t_2} X \rd t,
\]
where $t_{1,2}=m^2-s\,(1\pm\beta)/2$ with $\beta$ being the speed of a final
meson in the CMS given by 
\be
\label{beta}
\beta=\sqrt{1-\frac{4m^2}{s}}.
\ee
The result of the quadrature is
\[
Y=\frac{s\beta^3}{3m^4}\left(s^2+20m^2s+12m^4\right).
\]
Putting everything together, we obtain
\[
\sigma(s)=\frac{\pi\alpha^2\beta^3}{12s}\left(\frac{s^2}{m^4}+20\frac{s}{m^2}
+12\right)\left|
\frac{R_k}{s-M_k^2+\jj M_k\Gamma_k}
\right|^2,\nonumber
\]
where $\alpha$ is the fine-structure constant. Summing the $n$ diagrams of the 
type presented in Fig. \ref{fig:ee2vbarv} with various vector mesons $V_k$ 
coherently (allowing for additional phases $\delta_k$), and defining the 
coefficient
\be
\label{ck}
C_k=\frac{\pi\alpha^2}{12}R_k^2,
\ee
we arrive at the formula
\bea
\label{sigma}
\sigma(s)&=&\frac{\beta^3}{s}\left(\frac{s^2}{m^4}+20\frac{s}{m^2}
+12\right)\nl
&\times&
\left|\sum_{k=1}^n
\frac{\sqrt{C_k}\,e^{\jj\delta_k}}{s-M_k^2+\jj M_k\Gamma_k}
\right|^2.
\eea
The cross section here must be understood in the Born sense because we have 
used the bare photon propagator \cite{actis}.

The formula \rf{sigma} is valid for both the intermediate resonances and 
subthreshold poles. For a latter, the corresponding $\Gamma_k$ is zero.

For a resonance, we can proceed further and relate the coefficient $C_k$
to its decay properties. We first use the vertex \rf{vertex} 
to derive the formula
\be
\label{gammavv}
\Gamma_{V_k\ra\vv}=\frac{G^2_{V_kV\bar V}M_k\beta_k^3}{192\pi}
\left(\frac{M_k^4}{m^4}+20\frac{M_k^2}{m^2}+12\right),
\ee
where $\beta_k$ is given by Eq. \rf{beta} with $s=M_k^2$.

Then, taking into account the $\gamma V_k$ junction and again using the
bare photon propagator, we easily get for the Born $\die$ decay width the 
formula
\be
\label{gammaee}
\Gamma_{V_k\ra\die}=\frac{4\pi\alpha^2}{3}\frac{M_k}{g_k^2}.
\ee
Combining the relation \rf{rk}, \rf{ck}, \rf{gammavv}, and \rf{gammaee},
we arrive at 
\be
\label{ckfinal}
C_k=\frac{12\pi M_k^2\ \Gamma_{V_k\ra\vv}\,\Gamma_{V_k\ra\die}}
{\beta_k^3
\left(\frac{M_k^4}{m^4}+20\frac{M_k^2}{m^2}+12\right)}.
\ee
Formulas \rf{sigma} and \rf{ckfinal} are analogous to somewhat simpler ones 
from the standard BW formalism, which is valid for the $\die$ 
annihilation into two pseudoscalar mesons.

In the next section, we will explore the formula \rf{sigma} and treat $M_k$s, 
$\Gamma_k$s, $C_k$s, and $\delta_k$s as free parameters  (except $\delta_1$, 
which is kept at 0). We will determine their mean values and dispersions by 
fitting the experimental cross section using the standard $\chi^2$ criterion 
\cite{minuit}.

\section{Fitting the BESIII 2022 data}
The BESIII detector is situated at the symmetric $\die$ storage ring BEPCII at
the IHEP laboratory in Beijing, China. The collaboration working there
provided \cite{besiii2022} a very precise measurement of the $\die\to\dd$ 
Born cross section covering the CMS energy range from 4.08 to 4.60 GeV in 28 
points. High precision and point density enable testing of theoretical models.

We started fitting the experimental cross section using the formula 
\rf{sigma} with one resonance, expecting to find a resonance-like structure 
peaked at about 4.1~GeV or 4.4~GeV. As a result, we got a subthreshold pole
($M=3778\pm24$~MeV, $\Gamma=0\pm160$~MeV) instead, which generated a smooth
background, and a very bad $\cndf=2681/25$. 

Assuming two resonances resulted to $\cndf=217/21$ and the resonance 
parameters pointing to $\psi(4040)$ and $\psi(4230)$. The structure peaked 
at 4.1 GeV is an outcome of the interference of those two. 

\begin{figure}[]
\includegraphics[width=0.49\textwidth,height=0.49\textwidth]{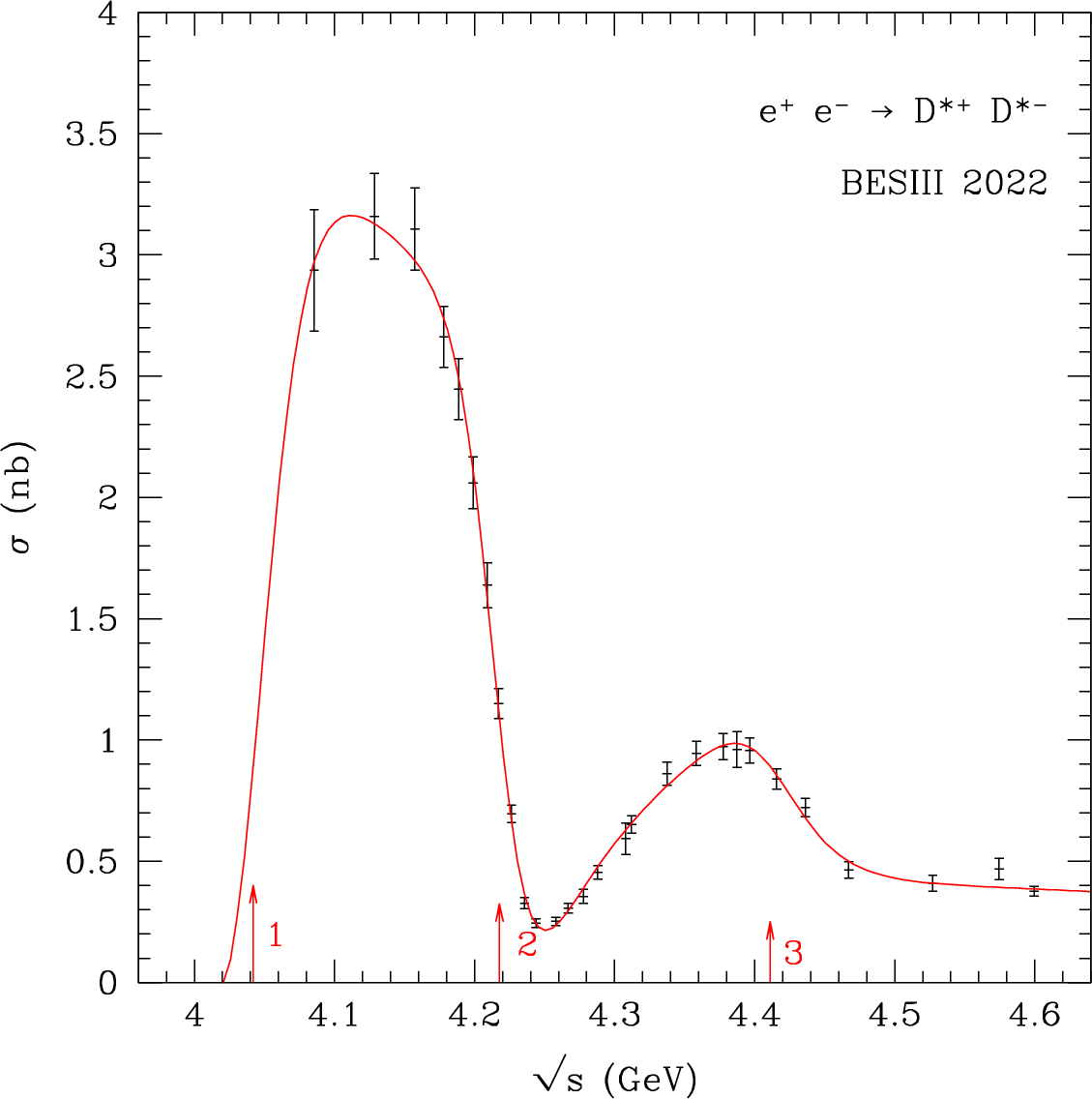}
\caption{\label{fig:3reso}Excitation function (red online) obtained as a fit 
of our model with three resonances to the BESIII data \cite{besiii2022}.
The arrows mark the positions of resonances, and their length is proportional
to local significance.}
\end{figure}

Finally, adding the third resonance to the fitting formula, we got an 
acceptable fit ($\cndf=15.0/17$, \p of 52\%), shown in Fig. \ref{fig:3reso}. 
Its parameters are summarized in Table \ref{tab:3reso}. 
There, we also show the local significances $\Sigma_k$ of resonances, estimated 
as ratios of $Q_k$'s mean values to their dispersions and the values of the
quantity
\be
\label{xk}
X_k=\Gamma_{V_k\to\dstp\dstm}\times\Gamma_{V_k\to\die}\,,
\ee
calculated from the obtained parameters by inverting the formula \rf{ck}.
\begin{table}[h]
\caption{\label{tab:3reso}Parameters of the three-resonance fit to the BESIII
2022 data \cite{besiii2022} based on Eq. \rf{sigma}. The $X_k$ is defined by
Eq. \rf{xk} and  the estimates of local significances are denoted as 
$\Sigma_k$.}
\begin{tabular*}{8.6cm}[b]{lccc}
\hline
\hline
\multicolumn{4}{c}
 {$\chi^2$/NDF=16.0/17~~~~~~~~~~\p = 52\%}         \\
\hline
$k$             &     1      &      2           & 3          \\
\hline
$M_k$ (MeV)&~~~~4042$\pm$20~~~&~~~4217.8$\pm$2.5~~~&~~~4411$\pm$14\\
$\Gamma_k$ (MeV)& 108$\pm$19 & 83.4$\pm$5.5      & 116$\pm$15\\
$C_k\times10^6$ (GeV$^4$) & 71$\pm$12   & 2.81(58)  &  4.4$\pm$1.2 \\
$\delta_k$ (rad)& 0 (f)      & -2.383(59)        & -2.299(45)\\
$X_k\times10^2$ (MeV$^2$)&$1.15\pm0.43$ & $1.38\pm0.28$ & $5.5\pm1.5$ \\
$\Sigma_k$    & 5.9\,$\sigma$ &4.8\,$\sigma$&  3.7\,$\sigma$ \\
\hline
\hline
\end{tabular*}
\end{table}

In an effort to find a better solution (in a $\chi^2$ sense), we repeated
the minimization procedure many times with randomly chosen starting
parameters. We found none. But we found several solutions with the same
$\chi^2$, $M_k$s, and $\Gamma_k$s \cite{multinote}, substantially differing in 
values of $C_k$s, $\delta_k$s, and, of course, of $X_k$s, and $\Sigma_k$s 
calculated from them. This phenomenon of multiple solutions is 
explained in the literature \cite{multisol}. It does not influence either
excitation function in Fig. \ref{fig:3reso}, or values of $\chi^2$, $M_k$s, 
and $\Gamma_k$s presented in Table \ref{tab:3reso}. However, the values of 
other quantities there must be taken with caution, as one of many possibilities. 
To pick up the right one, we would need to know the quantity $X_k$ \rf{xk} 
for at least one resonance.

\begin{figure}[b]
\includegraphics[width=0.49\textwidth,height=0.245\textwidth]{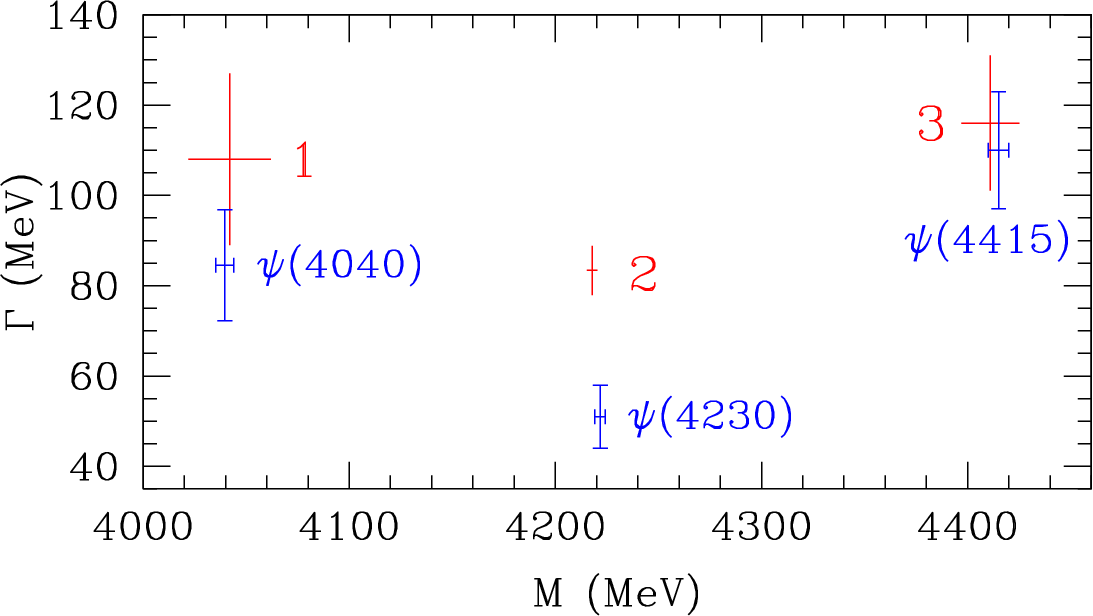}
\caption{\label{fig:mg_besiii}Comparison of the resonances found here
(marked with simple numerals, red online) with those in the PDG 2026 tables, 
distinguished by short lines at endpoints (blue online).}
\end{figure}

The comparison of masses and widths of the resonances we found with those 
listed in PDG tables \cite{pdg2026} is shown in Fig. \ref{fig:mg_besiii}. 

The central value of the $\psi(4040)$ mass agrees with the PDG value very well, 
but the error is much higher. Also, the width is off by one $\sigma$. These
deficiencies can be attributed to missing experimental data in the important
interval between the threshold and 4.0854 GeV (see Fig. \ref{fig:3reso}).

Also, the mass of $\psi(4230)$, which makes the steep downhill slope,
agrees well with PDG. But our width value, $83.4\pm5.5$~MeV, is significantly 
higher than the PDG's average value of $51\pm7$~MeV. Nevertheless, our value 
agrees
with those found in four recent BESIII experiments: $(80.7\pm4.4\pm1.4)$~MeV
\cite{24t}, $(71.7\pm16.2\pm32.8)$~MeV \cite{23u}, $(81.6\pm17.8\pm9.0)$~MeV
\cite{23x}, and $(72.9\pm6.1\pm30.8)$~MeV \cite{22au}. 

According to PDG, the $\dd$ decay of the $\psi(4230)$ has not yet been seen
observed, so this may be its first indication.

The parameters of $\psi(4415)$ agree with those of the PDG perfectly.

An interesting feature in Fig. \ref{fig:3reso} is that the
resonance-like structure with a peak at about 4.1 GeV is not a resonance,
but a result of the interference of two resonances, $\psi(4040)$ and
$\psi(4230)$.

\section{Conclusions}

We have used the simplest possible parametrization of the resonances' shape:
fixed masses $M_k$ and constant total widths $\Gamma_k$ and have gotten
acceptable results when fitting the $\die\ra\dd$ data by BESIII
Collaboration \cite{besiii2022}. 

In some cases, more sophisticated shape-line parametrizations may be needed 
in the basic formula \rf{sigma} to obtain good results. There are many 
possibilities of improving the resonances' description: energy-dependent widths 
$\Gamma_k(s)$, 
running masses $M_k(s)$ determined from $\Gamma_k(s)$ by the once- \cite{once} 
or twice- \cite{twice} subtracted dispersion relations, Flatt\'e approximation 
\cite{flatte}, modified Flatt\'e approximation \cite{mflatte}, $T$-matrix 
\cite{tmatrix}, and $K$-matrix \cite{kmatrix}.

To stress the necessity of using a correct BW-like formula for a particular 
final state, let us compare the behavior of formulas for (a) two 
pseudoscalar mesons, (b) a pseudoscalar and a vector meson, and (c) two 
vector mesons. Approaching the threshold, all three formulas
behave like $q^3$, where $q=|\vec k^\ast_{1,2}|$ is the magnitude of a final 
meson momentum in CMS. At asymptotically high energies ($s\gg
m^2$, $s\gg M_k^2$ for all $k$), their behaviours differ:
$\sigma_a\propto s^{-3}$, $\sigma_b\propto s^{-2}$, and 
$\sigma_c\propto s^{-1}$.

\begin{acknowledgments}
I thank Josef Jur\'{a}\v{n} for checking all formulas presented in this
Letter using Wolfram's \Mathematica~and for useful discussions.
\end{acknowledgments}

\end{document}